\input mtexsis
\paper
\singlespaced
\widenspacing
\twelvepoint
\Eurostyletrue
\sectionminspace=0.1\vsize
%definitions
\def\discr{{\bf S }_2} 
\def\parlam{^{\scriptstyle\lambda}}
\def\smallo{{\textstyle{1\over\sqrt{2}}}}
\def\nc{{N}_c}
\def\yo1{{{f_\pi}^2}}

\def\oneht{\textstyle{1\over 2} }
\def\onehtsq{\textstyle{1\over{\sqrt{2}}} }

\def\sss{\scriptscriptstyle}

\def\ql{{ q_{\sss L} }}
\def\qr{{ q_{\sss R} }}
\def\pl{{ p_{\sss L} }}
\def\pr{{ p_{\sss R} }}
\def\bql{{ {\bar q}_{\sss L} }}
\def\bqr{{ {\bar q}_{\sss R} }}
\def\bpl{{ {\bar p}_{\sss L} }}
\def\bpr{{ {\bar p}_{\sss R} }}
%references
\referencelist
\reference{pdg} Particle Data Group, \journal Phys. Rev.D;54, (1996)
\endreference
\reference{avw} C.~Lovelace, \journal Phys. Lett. B;28,264 (1968)
\endreference
\reference{*avwa}  M.~Ademollo, G.~Veneziano, and S.~Weinberg, 
                  \journal Phys. Rev. Lett.;22,83 (1969)
\endreference
\reference{lew} D.~Lewellen, \journal Nucl. Phys. B;392,137 (1993)
\endreference
\reference{*lewa} J.~Polchinski, UTTG-16-92, {\tt hep-th/9210045}
\endreference
\reference{alg}  S.~Weinberg, \journal Phys. Rev.;177,2604 (1969)
\endreference
\reference{*alga} See also, F.J.~Gilman and H.~Harari,
                  \journal Phys. Rev.;165,1803 (1968)
\endreference
\reference{*algb} B.~Zumino, in {\it Theory and Phenomenology in Particle
Physics}, 1968 International School of Physics `Ettore Majorana'
(Academic Press, New York, 1969) p.42
\endreference
\reference{*algc} S.~Weinberg, in {\it Lectures on Elementary Particles and
Quantum Field Theory}, edited by Stanley Deser {\it et al}, 
(MIT Press, Cambridge, MA, 1970) p.285
\endreference 
\reference{*algd}  R.~de Alfaro, S.~Fubini, G.~Furlan, and G.~Rossetti, 
            {\it Currents in Hadron Physics}, (North-Holland, Amsterdam, 1973)
\endreference
\reference{mended}  S.~Weinberg, \journal Phys. Rev. Lett.;65,1177 (1990);
           {\it ibid}, 1181
\endreference
\reference{bean} S.R.~Beane, DUKE-TH-95-98, {{\tt hep-ph/9512228}}
\endreference
\reference{*beana} See also, W.A.~Bardeen and C.T.~Hill,
\journal Phys. Rev. D;49,409 (1994)
\endreference
\reference{wilson}  K.G.~Wilson and D.G.~Robertson,  OSU-NT-94-08, {{\tt hep-th/9411007}}
\endreference
\reference{mustaki}  D.~Mustaki, {{\tt hep-ph/9410313}}
\endreference
\reference{mat}  M.~Burkardt, \journal Adv. Nucl. Phys.;23,1 (1996), {{\tt hep-ph/9505259}}
\endreference
\reference{suss}  A.~Casher and L.~Susskind,
\journal Phys. Lett. B;44,171 (1973);\hfill
\journal Phys. Rev. D;9,436 (1973)
\endreference
\reference{*sussa} See also, L.~Susskind and M.~Burkardt, {{\tt hep-ph/9410313}}
\endreference
\reference{sfsr}  S.~Weinberg, \journal Phys. Rev. Lett.;18,507 (1967)
\endreference
\reference{ksrf} K.~Kawarabayashi and M.~Suzuki,
                    \journal Phys. Rev. Lett.;38,883 (1966)
\endreference
\reference{*ksrfa} Riazuddin and Fayyazuddin,
                    \journal Phys. Rev.;147,1071 (1966)
\endreference
\reference{unbreak}  S.~Weinberg, in {\it Proceedings of Salam festschrift}, (Trieste, 1993), p.3
\endreference
\reference{beane} S.R.~Beane, work in progress
\endreference
\reference{krauss} See, for instance, L.M.~Krauss and F.~Wilczek, 
      \journal Phys. Rev. Lett.;62,1221 (1989)
\endreference
\reference{sei}  N.~Seiberg, 
\journal  Nucl. Phys. B;435,129 (1995), {{\tt hep-th/9411149}}
\endreference
\reference{*seia}  R.G.~Leigh and M.J.~Strassler, RU-96-101, {{\tt hep-th/9611020}}\endreference
\reference{*seib}  J.~Distler and A.~Karch, UTTG-20-96, {{\tt hep-th/9611088}}
\endreference
\reference{mandel} G.~'t Hooft, in {\it High Energy Physics}, 
Proceedings of the European Physical Society International Conference,
Palermo, 1975, ed.  A.~Zichichi (Editrice Compositori, Bologna, 1976),
p.1225; see also, S.~Mandelstam \journal Phys. Rev. D;19,2391 (1979)
\endreference
\endreferencelist
% title page
\titlepage
\obeylines
\hskip4.8in{DUKE-TH-96-121}
\hskip4.8in{DOE/ER/40762-107}
\hskip4.8in{U.ofMd.PP\#97-075}
\hskip4.8in{hep-ph/9706246}\unobeylines
%\line{ \hfill \TeXsis\ \fmtversion}     % a banner of sorts
\title
Flavor Doubling and 
the Nature of Asymptopia
\endtitle
\author
Silas R.~Beane

Department of Physics, Duke University
Durham, NC 27708-0305

{\it and}

Department of Physics, University of Maryland\Footnote\dag{Present Address.\hfill}
College Park, MD 20742-4111
\vskip0.1in
\center{{\it sbeane@pion.umd.edu}}\endcenter
\endauthor

\abstract
\singlespaced
\widenspacing

We consider the possibility that QCD with ${{N}}$ flavors has a useful
low-energy description with $2{{N}}$ flavors. Specifically, we
investigate a free theory of $2{{N}}$ quarks. Although the free theory
is $U({{{N}}})_L\times U({{{N}}})_R$ invariant, it admits a larger
$U(2{{{N}}})$ invariance. However, when the axial anomaly is accounted
for in the effective theory by a 't Hooft interaction, only
$SU({{{N}}})_L\times SU({{{N}}})_R\times U(1)_B \subset U(2{{{N}}})$
survives.  There is however a residual discrete symmetry that is not a
symmetry of the QCD lagrangian. This $\discr$ subgroup of
$U(2{{{N}}})$ has many interesting properties. For instance, when
explicit chiral symmetry breaking effects are present, $\discr$ is
broken unless ${\bar\theta}$=$0$ or $\pi$. By expressing the free
theory on the light-front, we show that flavor doubling implies
several superconvergence relations in pion-hadron scattering. Implicit
in the $2N$-flavor effective theory is a Regge trajectory with vacuum
quantum numbers and unit intercept whose behavior is constrained by
$\discr$.  In particular, $\discr$ implies that forward pion-hadron
scattering becomes purely elastic at high-energies, in good agreement
with experiment.

\endabstract 
\endtitlepage
\vfill\eject                                     % new page
% introduction
\superrefsfalse
\singlespaced
\widenspacing

%%%%%%%%%%%%%%%%%%%%%%%%%%%%%%%%%%%%%%%%%%%%%%%%%%%%%%%%%%%%%%%%%%%
\vskip0.1in
\noindent {\twelvepoint{\bf 1.\quad Introduction}}
\vskip0.1in

The hadron spectrum clearly exhibits regularity that is not explained
by symmetries of the QCD lagrangian.  For instance, to good
approximation observed mesons and baryons fall on linear Regge
trajectories with a universal slope parameter\ref{pdg}. Furthermore,
hadrons of different character exhibit universal mass-squared
splittings with remarkable accuracy\ref{avw}.  While QCD remains
intractable at low energies one might hope to find an effective
description which exhibits new regularity. One possibility
investigated by many is that QCD is in some sense equivalent to a
string theory. Presumably the world-sheet symmetries of an effective
string description could explain features of hadron phenomenology not
addressed by symmetries of the QCD lagrangian\ref{lew}. Although a
consistent stringy description has not been found, there is one
property of string models which certainly plays a fundamental and yet
ill understood role in hadron physics: the pattern of asymptotic
behavior of scattering amplitudes.

Asymptotic behavior in forward pion-hadron scattering is of special
interest. The Goldstone nature of the pion allows a precise
identification between Regge asymptotic behavior and algebraic
constraints on the hadronic mass-squared matrix\ref{alg}.  In
particular, superconvergent sum rules in pion-hadron scattering have a
one-to-one correspondence with symmetry properties of the hadronic
mass-squared matrix. These algebraic constraints have always appeared
mysterious, in part because they seem to have no place in the naive
quark model. In this paper, we will show that the mass-squared matrix
constraints and therefore the pattern of asymptotic behavior which
they imply, can be understood very simply as consequences of flavor
doubling.

One might object that such an underlying structure can have nothing to
do with the real world since the number of flavors is known. We
immediately reassure the reader that we are {\it not} suggesting a
modification of QCD, which in this paper has ${{N}}$ quark flavors.
Although one can count the number of QCD flavors and colors in the
low-energy theory through the effects of anomalies, it is important to
realize that as a matter of principle it is not possible to measure
the number of effective flavors inside a hadron at low energies, since
quarks are not gauge invariant objects. In practice, one infers the
number of effective flavors by observing global symmetries and their
corresponding conserved currents in low-energy
measurements. Therefore, in principle, there is no reason why QCD with
${{N}}$ flavors cannot have an effective low-energy description with
$2{{{N}}}$ flavors. What it is important to ask is whether it is
possible to double the number of flavors without introducing new
conserved currents, since such currents would almost certainly be in
violent disagreement with experiment.  We will show that doubling the
number of flavors introduces no new conserved currents when the axial
anomaly is taken into account in the effective theory.  However, we
will find that one cannot avoid a residual discrete symmetry which is
certainly {\it not} a symmetry of the QCD lagrangian.  Surprisingly,
this discrete symmetry has distinct consequences that are in agreement
with experiment.  For instance, there can be no P and CP violation in
the low-energy theory if this discrete symmetry is unbroken.
Furthermore, a constraint on the hadronic mass-squared matrix,
expressable as a superconvergent sum rule in pion-hadron scattering,
is a direct consequence of the discrete symmetry. This is one of the
algebraic sum rules alluded to above.

A further interesting property of the $2N$-flavor effective theory is
that it allows one to move from a current quark picture to a
constituent picture with massless and massive constituent quarks and a
conserved chiral current in the broken phase. Moreover, some of the
results suggested by the constituent quark picture can be verified
exactly using chiral symmetry and the new discrete symmetry in special
Lorentz frames.  Evidently doubling the number of flavors tells
something about aspects of low-energy phenomenology which are
mysterious from the point of view of the QCD lagrangian. At the end we
will entertain some speculations on why this is the case.

This paper is organized as follows.  In section 2 we review the
algebraic sum rules and their relation to Regge asymptotic behavior.
The main results of this section are summarized in table 1, which
provides our motivation for doubling the number of flavors.  In
section 3 we introduce a simple free-field model with $2{{{N}}}$
flavors and discuss its symmetries.  We investigate how this symmetry
structure is modified by a 't Hooft interaction. We then assume
dynamical chiral symmetry breaking, and develop a simple constituent
quark model. A conserved chiral current arises naturally in the broken
phase, leading to the conjecture that Goldstone bosons are bound
states of massless quarks.  In section 4 we express the free theory of
$2{{{N}}}$ flavors on the light-front, and show that the algebraic sum
rules of table 1 are fundamental properties of the effective
theory. In section 5 we use the symmetries of the $2N$-flavor
effective theory to find the chiral representations of the mesons, and
in turn prove that Goldstone bosons must be bound states of massless
constituent quarks as conjectured in section 3.  The relation between
the quark and meson mass-matrices is discussed in section 6.  In
section 7 we examine in some detail the phenomenology of the chiral
representation involving the pion.  We summarize our findings and
discuss their possible meaning in section 8.
%%%%%%%%%%%%%%%%%%%%%%%%%%%%%%%%%%%%%%%%%%%%%%%%%%%%%%%%%%%%%%%%%%
\vskip0.1in
\noindent {\twelvepoint{\bf 2.\quad Mysterious Regularity in the Hadron Spectrum}}
\vskip0.1in

Consider the process $\pi\alpha\rightarrow\pi\beta$, where $\alpha$
and $\beta$ are arbitrary single-hadron states, and $\pi$ is a
massless Goldstone boson.  Imagine writing down the most general
chiral invariant lagrangian involving all possible operators that can
contribute to this process.  There are an infinite number of operators
which couple the initial and final states to all possible intermediate
states in all channels, as well as an infinite number of contact
interactions.  Here we work in the approximation in which the
continuum is left out (the tree graph approximation). The discussion
is therefore exact for mesons in the large-$\nc$ limit\ref{mended}.

The existence of low-energy theorems, and more generally of chiral
perturbation theory, is a simple consequence of the fact that
Goldstone bosons couple only through derivative interactions. It is
less well known that this property of Goldstone bosons has additional
interesting implications.  Since the chiral lagrangian contains
operators with arbitrary numbers of derivatives, S-matrix elements
will have atrocious asymptotic behavior, unless there are intricate
cancellations among the various momentum-dependent operators. One can
obtain constraints based on the need for such cancellations by
considering an expansion in {\it inverse} energy in Lorentz frames
where all momenta are {\it collinear}\ref{alg}. We will discuss the
necessity of working in special Lorentz frames below.  The coefficient
in this expansion of zeroth-order in energy has special properties.
This coefficient contains the term which is protected by chiral
symmetry in the low-energy expansion, as well as contributions from
non-Goldstone intermediate particle states, and yet it contains no
unknown counterterms. All higher powers of energy contain unknown
counterterms and are therefore unconstrained by chirality. As in
\Ref{alg}, here we assume that these higher orders behave no
worse at high energies than the zeroth order term.  For ${{{N}}}=2$, the
zeroth-order coefficient takes the form\ref{alg}

$$\EQNalign{&{{C^{{(-)}\;\lambda}}_{\beta b,\alpha a}}\equiv
\lbrace i\epsilon _{abc}T_{c}- 
\lbrack {X_{a}\parlam},\, {X_{b}\parlam}\rbrack
\rbrace _{\beta\alpha}\EQN sc1;a \cr
&{{C^{{(+)}\;\lambda}}_{\beta b,\alpha a}}
\equiv\oneht\lbrace
\lbrack {X_{b}\parlam},\,\lbrack {X_{a}\parlam},\,{ M^{2}}
\rbrack\rbrack +
\lbrack {X_{a}\parlam},\,\lbrack {X_{b}\parlam},\,{ M^{2}}
\rbrack\rbrack\rbrace _{\beta\alpha} \EQN sc1;b \cr}$$
for the crossing-odd and crossing-even amplitudes, respectively. The
roman subscripts are isospin indices, ${T_a}$ are the isospin
matrices, and ${\hat M}^2$ is the hadronic mass-squared matrix.  The
helicity, $\lambda$, is a conserved quantum number in the collinear
frame, and $X_{a}^\lambda$ is an axial-vector coupling matrix, related
to the matrix element of the process ${\alpha (p,\lambda
)}\rightarrow{\beta (p',\lambda')}+{\pi}{(q,a)}$ in any frame in which
the momenta are collinear:

\offparens
$$ {{\cal M}_a}(p'{\lambda '}\beta ,p{\lambda }\alpha )=
{{({4{f_\pi}})}^{-1}} ({m_\alpha ^2}-{m_\beta ^2})\lbrack
{X_{a}\parlam}\rbrack _{\beta\alpha} {\delta _{{\lambda '}\lambda}}.
\EQN amp $$\autoparens
This identification holds both when $\alpha$ is at rest and moving at
infinite momentum\ref{alg}. The matrix $X^\lambda_a$ is independent of
the reference frame. Below, in dealing with quarks, we will relate to
this language by working in the light-cone frame.

It is important to realize that to this point we have assumed only
chirality and the tree graph approximation.  We will now assume that
the coefficients of \Eq{sc1} behave no worse at high energies than the
full amplitude is expected to behave on the basis of Regge
arguments\ref{alg}. Here by ``high energies'' we mean energies of
order the characteristic scale at which the momentum expansion
fails. It is conventional to denote this scale
$\Lambda_\chi\sim{m_\rho}$.  The crossing-odd amplitude is pure
${I_{t}=1}$, which Regge theory suggests is dominated by the $\rho$
trajectory, with intercept $\alpha_{1}(0)\simeq 0.5$.  $C^{(-)}$
vanishes if $\alpha_{1}(0) < 1$ and so one obtains the generalized
Adler-Weisberger (A-W) sum rule\ref{alg}

$$
{\lbrack {{X_{a}\parlam}},\,{{X_{b}\parlam}}\rbrack}_{\beta\alpha}
=i\epsilon _{abc}{(T_{c})}_{\beta\alpha}.
\EQN aw2 $$
Together with the defining relations, $\lbrack
T_{a},\,T_{b}\rbrack=i\epsilon_{abc}T_{c}$ and $\lbrack
T_{a},\,{{X_{b}\parlam}}\rbrack_{\beta\alpha} =
i\epsilon_{abc}{({X_{c}\parlam})}_{\beta\alpha}$, \Eq{aw2} closes the
chiral algebra.  It follows that for each helicity, $\lambda$, hadrons
fall into representations of $SU(2)\times SU(2)$, {\it in spite of the
fact that the group is spontaneously broken}\ref{alg}. However,
${X_a^\lambda}$ is not a true symmetry generator since it does not
commute with the mass-squared matrix, ${\hat M}^2$. As we will see, in
practice this means that the chiral representations in the broken
phase are reducible.

The crossing-even amplitude has both ${I_{t}=0}$ and ${I_{t}=2}$.
Since no ${I}=2$ states are observed in nature, Regge pole theory
suggests $\alpha_{2}(0)< 0$. The pomeron trajectory is expected to
dominate the ${I_{t}=0}$ channel and so one further expects
${\alpha_0}(0)=1$. Taken together these Regge constraints imply that
$C^{(+)}_{ba}$ is proportional to $\delta _{ba}$ (i.e. pure
${I_{t}}=0$).  It then follows that

\offparens
$$
{\lbrack {X_{b}\parlam},\,\lbrack { M^{2}},\,{X_{a}\parlam}
\rbrack\rbrack}_{\beta\alpha}\propto\delta_{ab}.
\EQN sc1 $$ \autoparens
This sum rule implies that the hadronic mass-squared matrix is the sum
of a chiral invariant and the fourth component of a chiral
four-vector\ref{alg}; that is,

\offparens
$$
{{\hat M}^2}={{\hat M}_0^2}(\lambda )
+{{\hat M}_{\sss \vev{{\bar q}q}}^2}(\lambda ),
\EQN sc1pure $$ \autoparens
in an obvious notation. Note that although the mass-squared matrix is
helicity independent, in principle the two parts of the mass-squared
matrix can be separately helicity dependent.  When ${{\hat M}_{\sss
\vev{{\bar q}q}}^2}(\lambda )$ vanishes, the mass matrix, ${{\hat
M}^2}$, commutes with ${X_a^\lambda}$, and hadrons of a given mass
form complete chiral multiplets.  One important consequence of
\Eq{aw2} and \Eq{sc1} is that hadrons which fall into an irreducible
representation must be degenerate\ref{alg}\ref{mended}.  A
particularly striking consequence of these algebraic sum rules is the
equation

\offparens
$$
{{C^{{(+)}\;\lambda}}_{\beta b,\alpha a}}=
-{\delta_{ab}}{{[{M_{\sss \vev{{\bar q}q}}^2}(\lambda )]}_{\beta\alpha}},
\EQN diffeqa$$\autoparens
which relates the crossing-even forward amplitude at high
energies to the symmetry breaking part of the mass-squared matrix.
This is a statement about diffraction: the constancy of cross sections
at high energies. Here diffraction, or rather the existence of a
pomeron, is equivalent to the existence of a non-vanishing chiral
order parameter\ref{alg}.

If ${{\hat M}_{\sss \vev{{\bar q}q}}^2}(\lambda )$ is diagonal,
scattering becomes purely elastic at high energies.  In Regge language
this constraint translates to ${\alpha_0} (0)<0$ for $\alpha\neq\beta$
--- no exchange of trajectories with vacuum quantum numbers when
$\alpha\neq\beta$. Algebraically, this superconvergence relation takes
the form

\offparens
$$[{{\hat M}_0^2}(\lambda ),
{{\hat M}_{\sss \vev{{\bar q}q}}^2}(\lambda )]=0.
\EQN mmconst2$$
\autoparens
This is perhaps the most puzzling of the sum rules since it implies a
successful phenomenology and yet appears to be a statement completely
unrelated to any QCD symmetry.  The phenomenological status of this
and the other sum rules is discussed in \Ref{alg}, \Ref{mended},
\Ref{bean} and below.  The algebraic sum rules have been generalized
to arbitrary numbers of flavors in \Ref{mended}.

\table{analog}
\caption{ 
The equivalence of the first and second columns is exact in the
tree-graph approximation (large-$\nc$ for pion-meson scattering). The
first row implies that, for each helicity, mass eigenstates fill out
generally reducible representations of $SU(2)_L\times SU(2)_R$.}
\doublespaced
\ruledtable
Hadrons   | Regge in $\pi\alpha\rightarrow\pi\beta$   \cr
$SU({2})_L\times SU({2})_R$|${\alpha_1} (0) < 1$ \cr
${{\hat M}^2}={{\hat M}_0^2}(\lambda )+{{\hat M}_{\sss \vev{{\bar q}q}}^2}(\lambda )$ |${\alpha_2} (0)<0$, \quad ${\alpha_0} (0)=1$\cr
$[{{\hat M}_0^2}(\lambda ),{{\hat M}_{\sss \vev{{\bar q}q}}^2}(\lambda )]=0$|${\alpha_0} (0)<0\quad \alpha\neq\beta$
\endruledtable
\endtable

In table 1 we summarize the algebraic constraints and their equivalent
statement in Regge theory.  As we will see, the Adler-Weisberger sum
rule, \Eq{aw2}, is not so mysterious; it is a straightforward
consequence of working in special Lorentz frames. On the other hand,
the constraints on the mass-squared matrix are puzzling, particularly
when viewed in the context of the quark model. We will discuss why
this is the case.  The primary purpose of this paper is to formulate
an underlying description in which the algebraic sum rules are
manifest.

%%%%%%%%%%%%%%%%%%%%%%%%%%%%%%%%%%%%%%%%%%%%%%%%%%%%%%%%%%%%%%%%%%%
\vskip0.1in
\noindent {\twelvepoint{\bf 3.\quad The q-p Model}}
\vskip0.1in

%%%%%%%%%%%%%%%%%%%%%%%%%%%%%%%%%%%%%%%%%%%%%%%%%%%%%%%%%%%%%%%%%%%
\vskip0.1in
\noindent {\twelvepoint{\it 3.1\quad Symmetries}}
\vskip0.1in

One way of constructing a field theory in which the algebraic sum
rules are manifest is to double the usual number of flavors.  In this
paper, we will investigate this possibility in detail.  The matter
content of our effective theory consists of $2N$ Dirac fermions
assembled into the vectors $q$ and $p$, each in the fundamental
representation of $SU({{{N}}})$, which transform with respect to
$SU({{{N}}})_L\times SU({{{N}}})_R$ as

\offparens
$$\eqalign{
&({{{N}}},1):\qquad\ql \rightarrow L\ql \qquad\pr \rightarrow L\pr \cr
&(1,{{{N}}}):\qquad \pl \rightarrow R\pl \qquad\qr \rightarrow R\qr .\cr}
\EQN qpal2$$
\autoparens
These $2N$ quarks are also assumed to carry a color charge, which will
be suppressed.  The most general $SU({{{N}}})_L\times SU({{{N}}})_R$
invariant free lagrangian one can build with $q$ and $p$ is

\offparens
$$
{\cal L}_0={\bar q}i\slashchar{\partial} q +{\bar p}i\slashchar{\partial} p
-{M_0}{\bql}\pr 
-{M_0'}{\bqr}\pl +h.c.
\EQN qpal3$$\autoparens
Parity conservation implies ${M_0}={M_0'}$ which gives

\table{symms}
\caption{Symmetries of the $q$-$p$ model. The left column lists the
familiar (classical) symmetries of the QCD lagrangian. The right
column exhibits the new symmetries arising from flavor
doubling. These symmetries continuously transform $q$ and $p$ into one
another, as is made clear by the presence of the off-diagonal Pauli
matrices, ${\sigma _{\sss 1}}$ and ${\sigma _{\sss 2}}$. The
$P$ subscript implies permutation of $q$ and $p$.}
\doublespaced
\ruledtable
${\delta\psi}/\psi$ | group \dbl ${\delta\psi}/\psi$ | group \CR
$-i\alpha$ | $U(1)_B$ \dbl $-i\phi{\sigma _{\sss 1}}$ | $U(1)_P$ \cr
$-i{{\vec\alpha}\cdot{\vec T}}$ | $SU({{{N}}})_V$ \dbl
$-i{{\vec\phi}\cdot{\vec T}}{\sigma _{\sss 1}}$ | $SU({{{N}}})_P$ \cr
$-i\beta{\sigma _{\sss 3}}{\gamma_5}$ | $U(1)_A$ \dbl $-i\omega{\sigma
_{\sss 2}}{\gamma_5}$ | $U(1)_{P5}$ \cr $-i{{\vec\beta}\cdot{\vec
T}}{\sigma _{\sss 3}}{\gamma_5}$ | $SU({{{N}}})_A$ \dbl
$-i{{\vec\omega}\cdot{\vec T}}{\sigma _{\sss 2}}{\gamma_5}$ |
$SU({{{N}}})_{P5}$
\endruledtable
\endtable

$$
{\cal L}_0={\bar q}i\slashchar{\partial} q +{\bar p}i\slashchar{\partial} p
-{M_0}({\bar q}p+{\bar p}q).
\EQN qpal3$$
This lagrangian clearly has symmetries beyond those assumed.  In order
to see the full symmetry structure it is convenient to define a new
field,

\offparens
$$\Psi=\onehtsq\left(\matrix{{q+p} \cr
                          {\gamma_5}({q-p})}\right). 
\EQN poiu$$\autoparens
In terms of this new field the lagrangian, \Eq{qpal3}, takes the familiar
form

$$
{\cal L}_0={\bar \Psi}i\slashchar{\partial} \Psi 
-{M_0}{\bar \Psi}\Psi .
\EQN qpal4$$
Since $\Psi$ is a $2N$-component vector, the full symmetry of the
lagrangian is $U(2{{{N}}})$. Note that in terms of this new field, we
need never mention chirality. For reasons that will become clear we
will show how $U(2{{{N}}})$ arises in the original basis. We can
assemble $q$ and $p$ into the $2N$-component vector: ${\psi}=({q}\;
{p})^{\sss T}$. The lagrangian,
\Eq{qpal3}, then takes the form

$$
{\cal L}_0={\bar \psi}i\slashchar{\partial} \psi 
-{M_0}{\bar \psi}{\sigma _{\sss 1}}\psi 
\EQN qpal4$$
where $\sigma _{\sss 1}$ is a Pauli matrix acting in the $q$-$p$
space. In this basis it is easy to classify symmetries.  In table 2 we
list the continuous global symmetries of \Eq{qpal4}. Note that
these are generally non-commuting symmetries. There are four
non-chiral symmetries given by $\delta\psi=-i\Theta\psi$ with
$\Theta=\lbrace \alpha,\, {{\vec\alpha}\cdot{\vec T}},\,
\phi{\sigma _{\sss 1}},\,
{{\vec\phi}\cdot{\vec T}}{\sigma _{\sss 1}}\rbrace$ which we denote
$U(1)_B$, $SU({{{N}}})_V$, $U(1)_P$ and $SU({{{N}}})_P$, respectively,
and four chiral symmetries given by
$\delta\psi=-i{\Theta_5}{\gamma_5}\psi$ with ${\Theta_5}=\lbrace
\omega{\sigma _{\sss 2}},\,
{{\vec\omega}\cdot{\vec T}}{\sigma _{\sss 2}},$ $\beta{\sigma _{\sss
3}},\, {{\vec\beta}\cdot{\vec T}}{\sigma _{\sss 3}}\rbrace$ which we
denote $U(1)_{P5}$, $SU({{{N}}})_{P5}$, $U(1)_{A}$ and
$SU({{{N}}})_{A}$, respectively. ${T_a}$ is an $SU({{{N}}})$ generator
and $\alpha$ ($\vec\alpha$), $\phi$ ($\vec\phi$), $\omega$
($\vec\omega$), and $\beta$ ($\vec\beta$) are arbitrary parameters
(${N^2}-1$ component vectors).  Of special interest is the $U(1)_P$
transformation

$$\psi\rightarrow e^{-i{\phi}{\sigma _{\sss 1}}}\psi .
\EQN disc$$
$U(1)_P$ is a subgroup of $U(2{{{N}}})$ which commutes with $\sigma
_{\sss 1}$.  Permutation of $q$ and $p$ generates the discrete
subgroup of $U(1)_{P}$ with $\phi =\pi/2$.  This subgroup is the group
${\bf S}_2$ of permutations of two objects$^1$\vfootnote1{ ${\bf S}_2$
is equivalent to ${\bf Z}_2$.}.  Hence, ${{\bf
S}_2}\subset{U(1)_{P}}\subset{U(2{{{N}}})}$.

It is straightforward to obtain the currents and conserved charges
associated with the symmetries of table 2. We define the $U(1)$
charges as $Q^{\sss B}$, ${\bar Q}^1\equiv{Q^{\sss P}}$, ${\bar
Q}^2\equiv{Q^{\sss P5}}$ and ${\bar Q}^3\equiv{Q^{\sss A}}$, and the
$SU({{{N}}})$ charges as ${Q_a^{\sss V}}$, ${G_a^1}\equiv{Q_a^{\sss
P}}$, ${G_a^2}\equiv{Q_a^{\sss P5}}$ and ${G_a^3}\equiv{Q_a^{\sss
A}}$.  It is then easy to show that these eight charges satisfy
the $U(2{{{N}}})$ algebra: 

\offparens
$$\EQNalign{&[{{\bar Q}^i},{{\bar Q}^j}]=i{\epsilon^{ijk}}{{\bar Q}^k} 
\qquad [{Q_a^{\sss V}},{Q_b^{\sss V}}]=i{f_{abc}}{Q_c^{\sss V}} 
\qquad [{{\bar Q}^i},{Q_a^{\sss V}}]=0 \EQN alg;a \cr
&[{G_a^i},{G_b^j}]=
{\textstyle{1\over{2{{{N}}}}}}
i{\epsilon^{ijk}}{\delta_{ab}}{{\bar Q}^k}+
{\textstyle{1\over 2}}
i{\epsilon^{ijk}}{d_{abc}}{G_c^k}+ 
{\textstyle{1\over 4}}
i{\delta^{ij}}{f_{abc}}{Q_c^{\sss V}} \EQN alg;b \cr
&[{{\bar Q}^i},{G_a^j} ]=i{\epsilon^{ijk}}{G_a^k} 
\qquad [{Q_a^{\sss V}},{G_b^i}]=i{f_{abc}}{G_c^i}
\EQN alg;c \cr
&[{Q^{\sss B}},{Q^{\sss B}}]=
[{Q^{\sss B}},{{\bar Q}^i}]=
[{Q^{\sss B}},{Q_a^{\sss V}}]=
[{Q^{\sss B}},{G_a^i}]=0, \EQN alg;d \cr}
$$\autoparens as expected. Here $\epsilon^{ijk}$ is the usual
antisymmetric $SU(2)$ tensor and $f_{abc}$ and $d_{abc}$ are
generalized Gell-Mann coefficients defined by $\lbrack
T_{a},T_{b}\rbrack=if_{abc}T_{c}$ and $\lbrace T_{a},T_{b}\rbrace={\bf
1} {\delta_{ab}}/{{{N}}}+ d_{abc}T_{c}$, where the $T_{a}$ are
$SU({{{N}}})$ generators normalized such that
$tr({T_{a}}{T_{b}})={\delta_{ab}}/2$.  The algebra of $U(2{{{N}}})$,
or $SU(2{{{N}}})\times U(1)_B$, arises from the embedding $SU(2)\times
SU({{{N}}})_V\rightarrow SU(2{{{N}}})$.  With ${{\bar\sigma} _{\sss
i}}\equiv{{\sigma} _{\sss i}}/2$, the generators in the defining
representation of $SU(2{{{N}}})$ can be written as $\{{{\bar\sigma}
_{\sss 1}}, {{\bar\sigma} _{\sss 2}}{\gamma_5}, {{\bar\sigma} _{\sss
3}}{\gamma_5}\}\otimes{\bf 1}$, ${\bf 1}\otimes{T_a}$ and
$\{{{\bar\sigma} _{\sss 1}},{{\bar\sigma} _{\sss
2}}{\gamma_5},{{\bar\sigma} _{\sss 3}}{\gamma_5}\}\otimes{T_a}$, which
are in correspondence with the charges ${\bar Q}^i$, ${Q_a^{\sss V}}$
and ${G_a^i}$, respectively.  This is similar to a spin-flavor
symmetry. Of course here the $SU(2)$ ---which mixes chiral and
non-chiral symmetries--- is a property of the special basis that we
have chosen.

Table 2 makes clear the motivation for working in the $q$-$p$ basis.
In this basis $U(2{{{N}}})$ is decomposed into subgroups which do not
mix $q$ and $p$ ---identified with classical QCD symmetries--- and
subgroups which mix $q$ and $p$.  Since we want to investigate the
relevance of flavor doubling to QCD, we must include an explicit
$U(1)_{A}$ violating ---$SU({{{N}}})_L\times SU({{{N}}})_R\times
U(1)_B$ preserving--- operator and therefore we must selectively break
$U(2{{{N}}})$.  Below we will see that explicit $U(1)_{A}$ breaking
effects also break $U(1)_{P5}$, $SU({{{N}}})_{P5}$ and
$SU({{{N}}})_{P}$ completely, and break $U(1)_{P}$ to its $\discr$
subgroup.

%%%%%%%%%%%%%%%%%%%%%%%%%%%%%%%%%%%%%%%%%%%%%%%%%%%%%%%%%%%%%%%%%%%
\vskip0.1in
\noindent {\twelvepoint{\it 3.2\quad The Effect of the Axial Anomaly}}
\vskip0.1in

We now add a simple $U(1)_A$ violating quark interaction to take into
account the effect of the axial anomaly.  This is a sensible thing to
do if we believe that the $2N$-flavor theory is a low-energy effective
theory of QCD. Consider the $U(1)_A$ violating, $SU({{{N}}})_L\times
SU({{{N}}})_R\times U(1)_B$ preserving 't Hooft interaction

\offparens
$$
{\cal L}''({\bar\theta} )=  
-{\kappa}\{ e^{i{\bar\theta}}\;det\; 
{\bar\psi}(1-{\sigma_{\sss 3}}{\gamma_5}){\psi} +
e^{-i{\bar\theta}}\;det\; 
{\bar\psi}(1+{\sigma_{\sss 3}}{\gamma_5}){\psi} \}, 
\EQN u1expbr$$\autoparens
where the determinant acts on $SU({{{N}}})$ matrices and $\kappa$ is a
new parameter of mass dimension $4-3{{{N}}}$. We have included a P and
CP violating phase, ${\bar\theta}$.  Note that ${\bpr}\pl +
{\bql}\qr =
\oneht {\bar\psi}(1-{\sigma_{\sss 3}}{\gamma_5}){\psi}$ 
and ${\bpl}\pr + {\bqr}\ql = 
\oneht {\bar\psi}(1+{\sigma_{\sss 3}}{\gamma_5}){\psi}$. 
One can verify that the 't Hooft interaction also breaks $U(1)_{P5}$,
$SU({{{N}}})_{P5}$ and $SU({{{N}}})_{P}$.  $U(1)_P$ is also
broken. However, a discrete subgroup survives; the $\discr$
transformation $\psi\rightarrow \pm i{\sigma _{\sss 1}}\psi$
interchanges ${\bar\psi}(1-{\sigma_{\sss 3}}{\gamma_5}){\psi}$ and
${\bar\psi}(1+{\sigma_{\sss 3}}{\gamma_5}){\psi}$, which is equivalent
to the transformation $L\leftrightarrow R$. That is,

\offparens
$$
{{\cal S}_{\sss 2}}\;{\cal L}''({\bar\theta} 
)\;{{\cal S}_{\sss 2}^{-1}}=
{\cal L}''(-{\bar\theta} ).
\EQN Pinvparoof$$\autoparens
Of course in the absence of explicit chiral symmetry breaking effects
we can perform the field redefinition, $\psi\rightarrow
e^{i{\bar\theta}{\sigma _{\sss 3}}{\gamma_5}/2N}\psi$, which removes
${\bar\theta}$ from the problem, and then P, CP {\it and} $\discr$ are
manifest discrete symmetries of the effective theory$^2$\vfootnote2{
Since $\psi\rightarrow\pm {\sigma _{\sss 1}}\psi$ is also an
invariance of the effective theory, strictly speaking the surviving
discrete symmetry is ${\bf Z}_4$ rather than ${{\bf Z}_2}=\discr$.}.

If we include an $\discr$ invariant explicit chiral symmetry breaking
term, $-{m_q}{\bar\psi}{\psi}$, we can again perform a field
redefinition which removes ${\bar\theta}$ from the 't Hooft
interaction.  For small ${\bar\theta}$, ${\bar\psi}{\psi}\rightarrow
{\bar\psi}{\psi} +i({\bar\theta}/N){\bar\psi}{\sigma_{\sss
3}}{\gamma_5}{\psi}$, which induces the P, CP and $\discr$ violating
operator,

\offparens
$$
-i{{m_q}\over N}\,{\bar\theta}\,{\bar\psi}{\sigma_{\sss 3}}{\gamma_5}{\psi}.
%-i({m_q}{\bar\theta}/N){\bar\psi}{\sigma_{\sss 3}}{\gamma_5}{\psi}.
\EQN ebs2$$\autoparens
Therefore, if $\discr$ is unbroken, the anomaly induces no P or CP
violation in the $2N$-flavor effective theory. In general, P, CP and
$\discr$ will be broken unless ${\bar\theta}$=$0$ or $\pi$.  Of
course, if $\discr$ is a global symmetry one does not expect it to be
exact and even soft-breaking on the scale of strong interaction
physics can grossly violate the experimental bound on $\bar\theta$.
We will discuss this point below.  In any case, given that
${\bar\theta}$ is an unconstrained parameter from the point of view of
QCD, one might conclude that the $2N$-flavor effective theory cannot
be describing the same low-energy physics as QCD. This might be the
case. However, as we will see, $\discr$ makes other distinct
predictions that agree well with experiment.

%%%%%%%%%%%%%%%%%%%%%%%%%%%%%%%%%%%%%%%%%%%%%%%%%%%%%%%%%%%%%%%%%%%
\vskip0.1in
\noindent {\twelvepoint{\it 3.3\quad The Constituent $q$-$p$ Model}}
\vskip0.1in

In summary, we have shown that because of the axial anomaly, doubling
the number of quark flavors does not introduce new conserved currents.
With the axial anomaly taken into account, our lagrangian takes the
form

$$
{\cal L}={\bar \psi}i\slashchar{\partial} \psi 
-{M_0}{\bar \psi}{\sigma _{\sss 1}}\psi 
-{\kappa}\{ det\; {\bar\psi}(1-{\sigma_{\sss 3}}{\gamma_5}){\psi}
+ det\; {\bar\psi}(1+{\sigma_{\sss 3}}{\gamma_5}){\psi} \}+\ldots
\EQN qeai$$
When $\kappa$=$0$, the lagrangian is $U(2{{{N}}})$ invariant.  When
$\kappa\neq 0$, $U(2{{{N}}})$ is broken explicitly to
$SU({{{N}}})_L\times SU({{{N}}})_R\times U(1)_B$ by the anomaly. The
lagrangian is also invariant with respect to the $\discr$
transformation $\psi\rightarrow{\sigma_{\sss 1}}\psi$. The dots refer
to other invariant operators. Here we are working in the chiral limit
and therefore P, CP and $\discr$ are exact symmetries of the theory.

We can now break $SU({{{N}}})_L\times SU({{{N}}})_R$ spontaneously 
to $SU({{{N}}})_V$ by assuming the non-vanishing condensates
 
$$
\vev{{\bar\psi}{\psi}}={v_{\sss 1}}
\qquad\vev{{\bar\psi}{\sigma_{\sss 3}}{\psi}}={v_{\sss 2}}.
\EQN mono1ba$$
Condensation of these quark bilinears is consistent with the QCD
pattern of chiral symmetry breaking$^3$\vfootnote3{Since the
't Hooft interaction is a $2N$-quark operator, the four-flavor
effective theory looks like an NJL model; presumably tuning $\kappa$
is one way of generating a condensate.}. If ${v_{\sss 2}}\neq 0$ then
$\discr$ is spontaneously broken. We will keep an open mind
regarding whether or not this is the case.  How do we learn about the
physical spectrum?  In particular, if the chiral symmetry is truly
spontaneously broken, where are the Goldstone bosons?  In the $q$-$p$
basis the most general free lagrangian consistent with parity is

$$
{\cal L}_0={\bar q}i\slashchar{\partial} q +{\bar p}i\slashchar{\partial} p
-{M_0}({\bar q}p+{\bar p}q)-{M_1}{\bar q}q-{M_2}{\bar p}p,
\EQN qpal3mod$$
where $M_1$ and $M_2$ are new undetermined parameters. It is important
to realize that $q$ and $p$ here are not the same as those above. The
original $q$ and $p$ are current quarks, whereas these are constituent
quarks. By inspection it is clear that if ${v_{\sss 2}}$=$0$, $\discr$
requires $M_1$=$M_2$.  It is useful to express $\discr$ invariance as
a general constraint on the mass matrix.  We can write the lagrangian,
\Eq{qpal3mod}, as

$$
{\cal L}_0={\bar \psi}i\slashchar{\partial} \psi -{\bar \psi}{\hat M }\psi ,
\EQN qpalgrapa$$
where, in an obvious notation, 

$$
{{\hat M}}={{\hat M}_0}+{{\hat M}_{\sss \vev{{\bar q}q}}},
\EQN grapa$$
with ${{\hat M}_0}$=${\sigma_{\sss 1}}{M_0}$ and ${{\hat M}_{\sss
\vev{{\bar q}q}}}$=${\bf 1}{({M_1}+{M_2})/2}+{\sigma_{\sss
3}}{({M_1}-{M_2})/2}$.  It is clear that ${\bar \psi}{{\hat M}_{\sss
\vev{{\bar q}q}}}\psi$ transforms like the condensates of
\Eq{mono1ba}. The action of $\discr$ on the lagrangian is such that

\offparens
$$
{{\cal S}_{\sss 2}}\;{\cal L}\;{{\cal S}_{\sss 2}^{-1}}-{\cal L}
\,\propto\,{\bar \psi}\, 
[{{\hat M}_0},{{\hat M}_{\sss \vev{{\bar q}q}}}]{\sigma _{\sss 1}}\,\psi . 
\EQN concurr$$\autoparens
So in order that $\discr$ be preserved,
it is sufficient that 

\offparens
$$
[{{\hat M}_0},{{\hat M}_{\sss \vev{{\bar q}q}}}]=0, 
\EQN uwo$$\autoparens
which clearly requires $M_1$=$M_2\equiv M$.  Note the similarity of
\Eq{grapa} and \Eq{uwo} to the algebraic sum rules of table 1.
Before addressing this issue in detail we can learn more about the
spectrum in the constituent quark model. For simplicity we will
assume throughout the rest of this section that ${v_2}=0$ and 
therefore $\discr$ is unbroken.

It is convenient to work in the diagonal
basis:

\offparens
$${\phi_\pm}\equiv{\textstyle {1\over\sqrt{2}}} (q\pm p).
\EQN physeig$$ 
\autoparens
These states transform as $({{{N}}} ,1 )\oplus (1,{{{N}}} )$, with
respect to $SU({{{N}}})_L\times SU({{{N}}})_R$ ---as deduced from
\Eq{qpal2}--- and as a doublet with respect to $\discr$.  In this basis, 
the lagrangian, \Eq{qpal3mod}, takes the form

\offparens
$$
{\cal L}_0={{\bar\phi}_+}i\slashchar{\partial}{\phi_+}
+{{\bar\phi}_-}i\slashchar{\partial}{\phi_-}
-(M+{M_0}){{\bar\phi}_+}{\phi_+} -(M-{M_0}){{\bar\phi}_-}{\phi_-}.
\EQN mono1a$$\autoparens
Are the Goldstone bosons implicit in this lagrangian?  If so, we
expect the presence of a conserved chiral current despite the fact
that the chiral symmetry is spontaneously broken. This is familiar
from current algebra.  This conserved current arises because it costs
the Goldstone bosons no energy to move from one point on the vacuum
manifold to another. Consider the axial vector current

$${A^\mu_a}={{\bar\phi}_-}{\gamma^\mu}{\gamma_5}{T_a}{\phi_-},
\EQN newchscurr$$
arising from the transformation
${\delta{\phi_-}}/{\phi_-}=-i{\gamma_5}{{\vec\omega}\cdot{\vec T}}$
where $\vec\omega$ is an ${N^2}-1$ component vector and $T_a$ is an
$SU({{{N}}})$ generator. This chiral transformation is not trivially
related to the original chiral symmetry.  One easily obtains

$${\partial_\mu}{A^\mu_a}={2i}{(M-{M_0})}{{\bar\phi}_-}{\gamma_5}{T_a}{\phi_-}.
\EQN divnewchscurr1$$
If we choose ${M}$=${M_0}$, the ${\phi_-}$ quarks are massless and we
have a conserved chiral current in the broken phase. This suggests
that ${M}$=${M_0}$ is a consequence of Goldstone's theorem and the
Goldstone modes are bound states of massless ${\phi_-}$ quarks with
interpolating field ${{\bar\phi}_-}{\gamma_5}{T_a}{\phi_-}$. For now
we will assume that this is the case. In a latter section we will use
chiral symmetry and $\discr$ to prove that the Goldstone bosons
must be ${\phi_-}$ bound states.  Note that the 't Hooft
operator contributes to the right side of
\Eq{divnewchscurr1}. Therefore, there must exist another distinct
operator which contributes with equal magnitude and opposite sign. Of
course consistency with Goldstone's theorem requires that the chiral
current of \Eq{newchscurr} remain conserved in the presence of
all interactions.

Finally, with ${M}$=${M_0}$ we find

\offparens
$$
{\cal L}_0={{\bar\phi}_+}i\slashchar{\partial}{\phi_+}
+{{\bar\phi}_-}i\slashchar{\partial}{\phi_-}
-2{M_0}{{\bar\phi}_+}{\phi_+}.
\EQN constmod$$\autoparens
In this constituent quark description, there are ${{N}}$ massless and
${{N}}$ massive quarks with a mass gap of $2{M_0}$. Because of symmetry
constraints the number of undetermined parameters has not changed.

It is straightforward to include explicit breaking effects.  Consider
the $\discr$ invariant current quark mass term

$$
{\cal L}'=-{m_q}({{\bar\phi}_+}{\phi_+}+{{\bar\phi}_-}{\phi_-})
=-{m_q}{\bar\psi}\psi .
\EQN mono1c$$
The constituent quark eigenvalues, including quark mass effects,
become

$$
{M_{\pm}}\equiv M\pm{M_0}+{m_q}.
\EQN masseswsb$$
The constituent quark masses therefore contain a piece that comes from
spontaneous chiral symmetry breaking, a piece which transforms as a
chiral singlet and a piece that transforms like the current quark
masses.  We then have

$${\partial_\mu}{A^\mu_a}={2i}{m_q}{{\bar\phi}_-}{\gamma_5}{T_a}{\phi_-},
\EQN divnewchscurr$$
when $M$=$M_0$, as expected, and the free constituent quark lagrangian becomes

\offparens
$$
{\cal L}_0 + {\cal L}'={{\bar\phi}_+}i\slashchar{\partial}{\phi_+}
+{{\bar\phi}_-}i\slashchar{\partial}{\phi_-}
-{m_q}{{\bar\phi}_-}{\phi_-}
-(2{M_0}+{m_q}){{\bar\phi}_+}{\phi_+}.
\EQN constmod$$\autoparens

%%%%%%%%%%%%%%%%%%%%%%%%%%%%%%%%%%%%%%%%%%%%%%%%%%%%%%%%%%%%%%%%%%%
\vskip0.1in
\noindent {\twelvepoint{\bf 4.\quad The $q$-$p$ Model on the Light-Front}}
\vskip0.1in

We saw in the last section that the properties of the mass matrix in
the constituent quark model resemble the algebraic sum rules in
pion-hadron scattering discussed in section 2. In this section we will
show that the algebraic relations of table 1 are fundamental
properties of the $2N$-flavor effective theory.  Since the algebraic
relations are derived in collinear frames where helicity is conserved,
in this section we will investigate the free effective theory in the
light-cone frame where helicity is also conserved, in order to make a
meaningful comparison.  Although we will not discuss the 't Hooft
interaction on the light-front, $U(1)_A$ and all symmetries which
continuously transform $q$ and $p$ into each other are assumed to be
anomalous. We will see that on the light-front the effective theory
with $2N$ flavors has special properties not shared by the naive quark
model.

The generalized Adler-Weisberger sum rule expresses the fact that for
each helicity hadron states fill out generally reducible
representations of the full chiral group in the broken phase.  The
specialization to helicity conserving Lorentz frames can be understood
as follows\ref{wilson}\ref{mustaki}\ref{mat}\ref{suss}.  In the
presence of the condensate the vacuum is teeming with quark-antiquark
pairs. Therefore, at rest the axial charges do not connect single
quark states but rather create quark-antiquark pairs. For this reason,
at rest the chiral algebra is not useful for classification purposes;
once the chiral symmetry is spontaneously broken there is no Lorentz
invariant sense in which hadrons fill out representations of the full
chiral group. On the other hand, if a system is moving past the vacuum
at infinite momentum there is a sense in which the vacuum, and
therefore the condensate, decouples from the system. Since helicity is
conserved in the infinite momentum frame, it should not be surprising
that ---for each helicity--- hadrons can be classified into
representations of the full chiral group. Mathematically, one can show
that on the light-front the axial charge operators conserve the number
of quarks and antiquarks separately and count the helicity of all the
quarks and antiquarks of a given state\ref{mustaki}.

One can show that the $2N$-flavor effective theory saturates the
Adler-Weisberger sum rules by expressing the free theory on the
light-front. The light-front Hamiltonian density corresponding to the
free-field Lagrangian given in \Eq{qpalgrapa} can be written as

\offparens
$$
{\cal H}=i{\sqrt{2}\over 4}\int d{y^{-}}\epsilon ({x^{-}}-{y^{-}})
{\psi^\dagger_+} ({y})
({{\hat M}^2}-{\Delta_\perp})
{\psi_+} ({x}),
\EQN lightfr1 $$\autoparens
where ${\psi_+}$ is the dynamical component of ${\psi}$ and $\epsilon
(x)$ is the sign function which satisfies ${\partial _x}{\epsilon
(x)}=2{\delta (x)}$. Here we assume that $[{{\hat M}},{T_a}]=0$.
Since the light-front vector current, ${{\tilde J}^{{\sss
V}\mu}_{a}}$, is trivially conserved, the Hamiltonian is
$SU({{{N}}})_L\times SU({{{N}}})_R$ invariant if the light-front axial
current

\offparens
$$
{{\tilde J}^{{\sss A}\mu}_{a}}={{J}^{{\sss A}\mu}_{a}}+
i{\textstyle {1\over 4}}{\bar \psi}{\gamma^\mu}{T_a}
{\gamma_5}\{{{\hat M}},{\sigma_{\sss 3}}\}
\int d{y^{-}}\epsilon ({x^{-}}-{y^{-}}){\gamma^+}{\psi_+} (y)
\EQN pregiardino $$\autoparens
is conserved. It is straightforward to find

\offparens
$$
{\partial_\mu}{{\tilde J}^{{\sss A}\mu}_{a}}
=\,-{\textstyle {1\over 4}}{\bar \psi}
{\gamma_5}[{{\hat M}^2},{\sigma_{\sss 3}}]{T_a}
\int d{y^{-}}\epsilon ({x^{-}}-{y^{-}}){\gamma^+}{\psi_+} (y).
\EQN giardino $$\autoparens
In the chiral invariant theory we have ${\hat M}$=${M_0}{\sigma _{\sss
1}}$ which gives ${{\hat M}^2}$=${M_0^2}$.  If we express the
Hamiltonian in the $q$-$p$ variables we see that it is diagonal in $q$
and $p$.

Consider $N=2$ QCD. We have the pattern of symmetry breaking
$SU(2)_L\times SU(2)_R\rightarrow SU(2)_V$. On the light-front this
breaking manifests itself through explicit breaking operators in the
Hamiltonian. In the broken phase, the axial charge operator is related
to the amplitude for the absorption and emission of
pions\ref{alg}\ref{mustaki}\ref{suss}.  Focusing on the two flavor
case, we can write

\offparens
$$
\eqalign{
&[{X_{a}^\lambda}]_{\beta\alpha}{\delta _{{\lambda '}\lambda}}=
\bra{\beta ,\lambda}{{\tilde Q}^{\sss A}_a}\ket{\alpha ,{\lambda '}} \cr
&{(T_{a})}_{\beta\alpha}=
\bra{\beta ,\lambda}{{\tilde Q}^{\sss V}_a}\ket{\alpha ,{\lambda '}}, \cr}
\EQN aw2whatever $$\autoparens
where ${\tilde Q}^{\sss V}_a$ and ${\tilde Q}^{\sss A}_a$ are the
light-front vector and axial-vector charges, respectively,
$X^\lambda_a$ is defined in \Eq{amp}, and we have made use of helicity
conservation.  Taking matrix elements of the light-front charge
algebra and inserting a complete set of states yields

$$
{\lbrack {{X_{a}\parlam}},\,{{X_{b}\parlam}}\rbrack}_{\beta\alpha}
=i\epsilon _{abc}{(T_{c})}_{\beta\alpha},
\EQN aw4 $$
which is of course the generalized Adler-Weisberger sum rule.  This
result is not surprising since on the light-front the four-flavor
effective theory is simply two copies of the free quark model, each of
which satisfies the sum rules\ref{mustaki}. However, the manifestation
of chiral symmetry breaking in the two pictures is fundamentally
different.

In the naive quark model, quark mass effects do not break chiral
symmetry on the light-front\ref{wilson}\ref{mustaki}. On the other
hand, in the $2N$-flavor effective theory, quark mass terms break
chiral symmetry both at rest and on the light-front.  This is easy to
see by performing a simple spurion analysis. We can assign mass terms
the $SU({{{N}}})_L\times SU({{{N}}})_R$ transformation properties

\offparens
$$
\eqalign{
&{M_0}\rightarrow {M_0}  \cr
&{\Gamma}\rightarrow L{\Gamma}{R^\dagger}=R{\Gamma}{L^\dagger}, \cr}
\EQN spurtran $$\autoparens
where ${\Gamma}=\{{m_q}, {M_1}, {M_2}\}$, and build all invariant
light-front bilinears.  Out of $M_0$ alone we can construct
${q_{+}^\dagger}{M_0^2}{q_{+}}$ and ${p_{+}^\dagger}{M_0^2}{p_{+}}$,
which of course appear in \Eq{lightfr1} with ${{\hat M}^2}$=${M_0^2}$.
If we allow a non-zero $\Gamma$, we have the additional chiral
invariant operators ${q_{+}^\dagger}{\Gamma ^2} {q_{+}}$ and
${p_{+}^\dagger}{\Gamma ^2} {p_{+}}$, as well as the operators
${q_{+}^\dagger}{\Gamma {M_0}} {p_{+}}$ and ${p_{+}^\dagger}{\Gamma
{M_0}} {q_{+}}$ {\it which break chiral symmetry explicitly}. With
$N=2$ these operators transform like the fourth component of a chiral
four-vector.  If we include explicit breaking due to current quark
masses, our mass-matrix becomes ${\hat M}$=${M_0}{\sigma _{\sss
1}}+{m_q}$ and so ${\hat M}^2$=${M_0^2}+{m_q^2}+2{m_q}{M_0}{\sigma
_{\sss 1}}$.  As is clear from the spurion analysis and \Eq{giardino},
the piece proportional to ${\sigma _{\sss 1}}$ breaks chiral symmetry
explicitly via the terms ${q_{+}^\dagger}{{m_q}{M_0}} {p_{+}}$ and
${p_{+}^\dagger}{{m_q}{M_0}} {q_{+}}$.  Therefore in the $2N$-flavor
effective theory, quark masses break chiral symmetry on the
light-front.

The way in which explicit chiral symmetry breaking enters in the quark
model is peculiar.  One can prove that on the light-front, one-body
operators that break chiral symmetry like the fourth component of a
chiral four-vector do not commute with the angular momentum
operator\ref{suss}. Such operators therefore break chiral symmetry
{\it and} spatial rotations.  As pointed out in
\Ref{suss}, one can get around this constraint by (i) introducing
spin-orbit couplings, or (ii) adding mirror fermions. It is very
interesting that several of the results we will derive below in the
four-flavor effective theory can also be obtained in a two-flavor
quark model by introducing operators with spin-orbit
couplings\ref{suss}.

In the light-front $q$-$p$ basis we can express the mass-squared
matrix as ${{\hat M}^2}={{\hat M}_0^2}+{{\hat M}_{\sss \vev{{\bar
q}q}}^2}$ where now

\offparens
$$
\eqalign{
&{{\hat M}_0^2}={\bf 1}\{ {M_0^2}+({M_1^{\prime 2}+M_2^{\prime 2}})/2 \}
+{\sigma_{\sss 3}}(M_1^{\prime 2}-M_2^{\prime 2})/2 \cr
&{{\hat M}_{\sss \vev{{\bar q}q}}^2}=
{\sigma_{\sss 1}}{M_0}({M'_1}+{M'_2})\cr}
\EQN iran $$\autoparens
with ${M'_i}\equiv{M_i}+{m_q}$.
Note that with
${\hat M}$, ${\hat M}_0$, and ${\hat M}_{\sss \vev{{\bar q}q}}$
defined in \Eq{grapa},  
${{\hat M}^2}$=${({{\hat M}})^2}$. However,
${{\hat M}_0^2}\neq {({{\hat M}_0})^2}$ and ${{\hat M}_{\sss
\vev{{\bar q}q}}^2}\neq {({{\hat M}_{\sss \vev{{\bar q}q}}})^2}$.
The chiral decomposition of ${{\hat M}^2}$ has been deduced from the
spurion transformation properties.

The action of $\discr$ on the light-front Hamiltonian is such that

\offparens
$$
{{\cal S}_{\sss 2}}\;{\cal H}\;{{\cal S}_{\sss 2}^{-1}}-{\cal H}
\,\propto\,
\int d{y^{-}}\epsilon ({x^{-}}-{y^{-}}){\psi^\dagger_+} ({y})
[{{\hat M}_0^2},{{\hat M}_{\sss \vev{{\bar q}q}}^2}]{\sigma _{\sss 1}}\,
{\psi_+} ({x}),
\EQN concurr$$\autoparens
where we have used \Eq{iran}. Therefore, $\discr$ is preserved if
$[{{\hat M}_0^2},{{\hat M}_{\sss
\vev{{\bar q}q}}^2}]=0$ which requires ${M_1}$=${M_2}$, as expected.
In this case, the symmetries of the mass-squared matrix can 
be summarized in the equations

\offparens
$$\EQNalign{
&{{\hat M}^2}={{\hat M}_0^2}+{{\hat M}_{\sss \vev{{\bar q}q}}^2}\EQN summ;a \cr
&\,\; [{{\hat M}_0^2},{{\hat M}_{\sss \vev{{\bar q}q}}^2}]=0. \EQN summ;b \cr}
$$\autoparens Together with \Eq{aw4}, these relations are familiar
from table 1 as algebraic sum rules in forward pion-hadron scattering.
Of course, here we have studied the symmetry properties of the quark
mass-squared matrix whereas the sum rules of table 1 are for the
hadronic mass-squared matrix.  However, since here the algebraic sum
rules are all statements of symmetry, they should hold also for the
quark bound states.  We will see that this is the case.  As shown in
section 2, \Eq{summ;a} implies the existence of a Regge trajectory
with vacuum quantum numbers and unit intercept; i.e. a pomeron. Here
${{\hat M}_{\sss
\vev{{\bar q}q}}^2}$ is independent of helicity. Therefore,
\Eq{diffeqa} implies that pion-hadron scattering in the effective theory is
helicity-independent at high-energies. Evidently, $\discr$ constrains
the pomeron trajectory since \Eq{summ;b} implies that forward
pion-hadron scattering becomes purely elastic at high energies.

We have now found two low-energy predictions of unbroken $\discr$
invariance: the absence of P and CP violation and the absence of
inelastic diffraction, both of which are consistent with nature and
yet neither of which is explained by QCD.  We can now use $\discr$
to learn more about hadron spectra in the effective theory.

%%%%%%%%%%%%%%%%%%%%%%%%%%%%%%%%%%%%%%%%%%%%%%%%%%%%%%%%%%%%%%%%%%%
\vskip0.1in
\noindent {\twelvepoint{\bf 5.\quad The Relevance of the Permutation Group}}
\vskip0.1in

%%%%%%%%%%%%%%%%%%%%%%%%%%%%%%%%%%%%%%%%%%%%%%%%%%%%%%%%%%%%%%%%%%%
\vskip0.1in
\noindent {\twelvepoint{\it 5.1\quad Symmetry Argument}}
\vskip0.1in

In this section we will discuss what can be learned about meson states
in the $2N$-flavor effective theory purely from symmetry
considerations.  Our symmetry breaking pattern is $G\rightarrow H$
where $G=SU({{{N}}})_L\times SU({{{N}}})_R$ and $H=SU({{{N}}})_V$. In
the broken phase, the physical hadron states $P_i$ fill out
irreducible representations of $H$.  Like the constituents $\phi_+$
and $\phi_-$ ---for each helicity--- the $P_i$ fall into generally
reducible representations of $G$. That is, the $P_i$ can be expressed
as linear combinations of states $B_i$ which are in irreducible
representations of $G$. A given reducible multiplet then takes the
general form:

\offparens
$$
\ket{P_{\sss 1}}= {\sum_{k=1}^n}{u_{\sss 1k}}\ket{B_k} 
\quad\cdots\quad
\ket{P_{\sss n}}= {\sum_{k=1}^n}{u_{\sss nk}}\ket{B_k}
\EQN genstates $$\autoparens
where the $u$'s are mixing angles. Since the mixing angles are not
fixed by $G$ and $n$ is arbitrary, {\it a priori} these
representations can be very complicated.  Say we restrict ourselves to
a subset of states within a given chiral multiplet that carry the same
$H$ charge. When the underlying theory is $\discr$ invariant, the
constituents $\phi_+$ and $\phi_-$ transform as an $\discr$
doublet. Therefore, we can require that the physical states of a given
$H$ charge be invariant with respect to arbitrary $\discr$
transformations:

$$\pmatrix{{B_i} \cr {B_j} \cr}
\rightarrow
\left(\matrix{0&1\cr
              1&0\cr}\right) 
\pmatrix{{B_i} \cr {B_j} \cr}\qquad \forall\; i,j .
\EQN matrix2$$ 
Although this might seem ad hoc, we will see below that this
transformation can be directly related to permutations of the
composites.  Invariant physical states must be either completely
symmetric or completely antisymmetric with respect to this
transformation.  There is a single completely symmetric state for any
$n$. However, there is a completely antisymmetric state only for
$n$=$2$.  There is therefore a single solution consistent with the
permutation symmetry:

\offparens
$$
\ket{P_{\sss 1}}={\textstyle {1\over\sqrt{2}}}\{\ket{B_{\sss 1}} 
+\ket{B_{\sss 2}} \} \qquad
\ket{P_{\sss 2}}={\textstyle {1\over\sqrt{2}}}\{\ket{B_{\sss 1}} 
-\ket{B_{\sss 2}} \}.
\EQN gensol $$\autoparens
This has been a naive way of finding that the only non-trivial
representation of $\discr$ is a doublet consisting of a symmetric
state and an antisymmetric state.  Hence, in general, we expect that a
physical hadron state will be either in an irreducible chiral
representation or in a reducible representation of the form
\Eq{gensol}, where the states of definite chirality contribute with
equal weight.  This multiplet structure has an immediate consequence.
Mass-squared splitting between the physical states comes about if the
matrix element $\bra{B_1}{{\hat M}^2}\ket{B_2}$ is non-vanishing.
This is so if ${\hat M}^2$ contains a piece which mixes different $G$
representations. We know that 
the full mass-squared matrix can be written ${\hat M}^2$=${\hat
M}^2_0$+${{\hat M}^2_{\sss
\vev{{\bar q} q}}}$, where ${\hat M}^2_0$ transforms like a $G$
singlet, and ${{\hat M}^2_{\sss \vev{{\bar q} q}}}$ breaks chiral
symmetry and therefore mixes different $G$ representations.  One can
then readily check that
\Eq{gensol} implies $\bra{P_1}{{\hat M}^2_0}\ket{P_2}$=
$\bra{P_1}{{\hat M}^2_{\sss \vev{{\bar q}q}}}\ket{P_2}$=$0$,
which immediately gives \Eq{summ;b}, as expected.

%%%%%%%%%%%%%%%%%%%%%%%%%%%%%%%%%%%%%%%%%%%%%%%%%%%%%%%%%%%%%%%%%%%
\vskip0.1in
\noindent {\twelvepoint{\it 5.2\quad ${{N}}$=$2$ Meson Multiplet Structure}}
\vskip0.1in

Now we apply our result to the two flavor case. We have the pattern of
symmetry breaking $SU(2)_L\times SU(2)_R\rightarrow
SU(2)_V$. Therefore, we expect physical hadron states to be states of
definite isospin and ---for each helicity--- to fill out generally
reducible representations of $SU(2)_L\times SU(2)_R$. All mass-squared
splittings transform in the $(2,2)$ (four-vector) representation.  The
quarks transform as $(1,2)$ or $(2,1)$ with respect to $SU(2)_L\times
SU(2)_R$. In what follows, $\mu^2$ and $\delta$ represent generic
elements of the mass-squared matrix which transform as $(1,1)$ and
$(2,2)$, respectively; that is, ${\mu^2} \in {{\hat M}_0^2}$ and
$\delta \in {{\hat M}_{\sss \vev{{\bar q}q}}^2}$.  Mesons states carry
isospin $0$ and $1$ and therefore transform as $(2,2)$, $(3,1)$,
$(1,3)$ and $(1,1)$.  One can easily check that the only products of
these states that contain four-vectors are $({2},{2})\otimes(3,1)$,
$({2},{2})\otimes(1,3)$, and $({2},{2})\otimes(1,1)$.

Physical meson states are states of definite parity and isospin and
are therefore, without loss of generality, can be considered
combinations of the isovectors ${\ket{2,2}_a}$,
${\ket{1,3}_a}-{\ket{3,1}_a}\equiv{\textstyle{\sqrt{2}}}{\ket{V}_a}$
and
${\ket{1,3}_a}+{\ket{3,1}_a}\equiv{\textstyle{\sqrt{2}}}{\ket{A}_a}$,
and the isoscalars ${\ket{2,2}_4}$ and $\ket{1,1}$. The subscripts are
isospin indices.  Mesons can fall into irreducible
representations. However all states within an irreducible
representation must be degenerate\ref{mended}.  Charge conjugation
leaves $(2,2)$ and $(1,1)$ unchanged and interchanges $(1,3)$ and
$(3,1)$\ref{alg}. Therefore only ${\ket{V}_a}$ changes sign under
charge conjugation.  The two simplest solutions consistent with the
multiplet structure implied by $\discr$ are:

\vskip0.1in
\noindent{\bf\underbar{Case a}}:\qquad 
$({2},{2})\otimes(3,1)\;{\it and}\; ({2},{2})\otimes(1,3)$
\vskip0.1in

\offparens
$$\eqalign{ 
&\ket{{\rm I}}_a=\smallo \{{\ket{2,2}_a}-{\ket{A}_a}\}
\hskip1.64in
{M_{\sss {I}}^2}={\mu^2}-\delta\cr
&\ket{{\rm II}}_a=\smallo \{{\ket{2,2}_a}+{\ket{A}_a}\}
\hskip1.575in
{M_{\sss {{II}}}^2}={\mu^2}+\delta\cr
&\ket{{\rm III}}={\ket{2,2}_4}\quad \ket{{\rm IV}}_a={\ket{V}_a}
\hskip1.404in
{M_{\sss III}^2}={M_{\sss {{IV}}}^2}={\mu^2}\cr
&\hskip3in{M_{\sss {I}}^2}+
{M_{\sss {{II}}}^2}=
2{M_{\sss III}^2}=2{M_{\sss {{IV}}}^2}.\cr}
\EQN mes1$$ 
\autoparens
States $\ket{\rm I}$, $\ket{\rm II}$ and $\ket{\rm III}$ have charge
conjugation sign $\pm\epsilon$, $\ket{\rm IV}$ has sign $\mp\epsilon$.
States $\ket{\rm I}$ and $\ket{\rm II}$ form an $\discr$ doublet.
In the right column we exhibit the mass relations implied by the
representation content. The lowest lying member of this quartet must
be an isovector.

\vskip0.1in
\noindent{\bf\underbar{Case b}}:\qquad $({2},{2})\otimes(1,1)$
\vskip0.1in

\offparens
$$\eqalign{ 
&\ket{{\rm I}}=\smallo \{{\ket{2,2}_4}-\ket{1,1}\}
\hskip2.025in {M_{\sss I}^2}={\mu^2}-\delta\cr
&\ket{{\rm II}}=\smallo \{{\ket{2,2}_4}+\ket{1,1}\}
\hskip1.95in {M_{\sss {II}}^2}={\mu^2}+\delta\cr
&\ket{{\rm III}}_a={\ket{2,2}_a}
\hskip2.73in{M_{\sss {{III}}}^2}={\mu^2}\cr
&\hskip3.55in
{M_{\sss I}^2}+{M_{\sss II}^2}=2{M_{\sss {{III}}}^2}\cr}
\EQN mes2$$ 
\autoparens
These states have the same charge conjugation sign. 
Again states $\ket{\rm I}$ and $\ket{\rm II}$ form an $\discr$ doublet.
The lowest lying member of this triplet must be an isoscalar.

We can also build other reducible representations which are consistent
with the $\discr$ symmetry. However, if we assume that our meson
states are bound states of two quarks, then the singlet $(1,1)$ can
only arise from the product $(1,2)\otimes(1,2)=(1,1\oplus 3)$, and
therefore the singlet $(1,1)$ is always grouped with the adjoint
$(1,3)$. This means that the singlet and the adjoint will never occur
as distinct states within the same chiral multiplet. Cases (a) and (b)
are then the only two possibilities.  We will see how this comes about
in more detail below when we build meson states out of constituent
quarks. This pairing of singlet and adjoint can also be thought a
consequence of the large-$\nc$ approximation\ref{mended}.  Of course
in the large-$\nc$ limit one also has singlet-adjoint degeneracy.

This multiplet structure is not new. These results have been derived
previously by assuming Regge behavior in pion-hadron scattering and
working directly with the algebraic constraints of table
1\ref{mended}.  In that derivation the algebraic nature of the sum
rules was somewhat mysterious. In the effective theory, these results
are consequences of chiral symmetry and the $\discr$ symmetry, with no
need for further assumptions. This leads to a powerful statement about
the ground state of the $2N$-flavor effective theory: {\it Since the
Goldstone bosons are isovector, they must fall into a representation
of type (a), and so must belong to a state of type $\ket{I}$}.  We can
now see if our identification of the Goldstone bosons as ${\phi_-}$
bound states is correct.
%%%%%%%%%%%%%%%%%%%%%%%%%%%%%%%%%%%%%%%%%%%%%%%%%%%%%%%%%%%%%%%%%%%
\vskip0.1in
\noindent {\twelvepoint{\it 5.3\quad Direct Construction}}
\vskip0.1in

Here we construct the meson states directly in the four-flavor
effective theory and compare with what we found purely on the basis of
symmetry.  Consider products of the quark states defined in
\Eq{physeig}:

\offparens
$$\eqalign{
&{{\bar\phi}^{\sss 1}_-}{\phi^{\sss 2}_-}\propto
({{\bar q}_{\sss 1}}{q_{\sss 2}}+{{\bar p}_{\sss 1}}{p_{\sss 2}})-
({{\bar q}_{\sss 1}}{p_{\sss 2}}+{{\bar p}_{\sss 1}}{q_{\sss 2}})\cr
&\quad\qquad\qquad(2,2)\qquad (1,1\oplus 3)\oplus (1\oplus 3,1)\cr
&{{\bar\phi}^{\sss 1}_+}{\phi^{\sss 2}_+}\propto
({{\bar q}_{\sss 1}}{q_{\sss 2}}+{{\bar p}_{\sss 1}}{p_{\sss 2}})+
({{\bar q}_{\sss 1}}{p_{\sss 2}}+{{\bar p}_{\sss 1}}{q_{\sss 2}})\cr
&\quad\qquad\qquad(2,2)\qquad (1,1\oplus 3)\oplus (1\oplus 3,1)\cr
&{{\bar\phi}^{\sss 1}_+}{\phi^{\sss 2}_-}\propto
({{\bar q}_{\sss 1}}{q_{\sss 2}}-{{\bar p}_{\sss 1}}{p_{\sss 2}})+
({{\bar p}_{\sss 1}}{q_{\sss 2}}-{{\bar q}_{\sss 1}}{p_{\sss 2}})\cr
&\quad\qquad\qquad(2,2)\qquad (1,1\oplus 3)\oplus (1\oplus 3,1)\cr
&{{\bar\phi}^{\sss 1}_-}{\phi^{\sss 2}_+}\propto
({{\bar q}_{\sss 1}}{q_{\sss 2}}-{{\bar p}_{\sss 1}}{p_{\sss 2}})-
({{\bar p}_{\sss 1}}{q_{\sss 2}}-{{\bar q}_{\sss 1}}{p_{\sss 2}})\cr
&\quad\qquad\qquad(2,2)\qquad (1,1\oplus 3)\oplus (1\oplus 3,1).\cr }
\EQN phenwhatpre$$\autoparens
The numerical scripts make the permutation properties clear, and we
have used the chiral transformation properties of $q$ and $p$ given in
\Eq{qpal2}. It is assumed that these are states of definite helicity,
parity and isospin. Note that the insertion of additional gamma
matrices can only change the parity of the state, or interchange the
$(2,2)$ and $(1,1\oplus 3)\oplus (1\oplus 3,1)$ representations.  Up
to a phase, these ``wavefunctions'' are invariant with respect to the
independent $\discr$ transformations

$$\EQNalign{& {q_{\sss 1}}\longleftrightarrow{p_{\sss 1}}\quad  
{q_{\sss 2}}\longleftrightarrow{p_{\sss 2}} \EQN cons;a \cr
& {q_{\sss i}}\longleftrightarrow{p_{\sss i}}\quad\;  
{q_{\sss j}},{p_{\sss j}}\quad
{\tt fixed}\;\quad{i\neq j}. \EQN cons;b \cr}
$$
We see explicitly that the product of two $\discr$ doublets gives
two $\discr$ doublets; i.e. $2\otimes 2 =2\oplus 2$. However,
invariance under charge conjugation necessarily unfolds one of the
doublets since ${{\bar\phi}_+}{\phi_-}$ and ${{\bar\phi}_-}{\phi_+}$
are not states of definite charge conjugation sign.  The composite
wavefunctions of definite charge conjugation sign and their associated
chiral representation content are:

$$\eqalign{
{\ket{\rm I}}&\sim
{{\bar\phi}^{\sss 1}_-}{\phi^{\sss 2}_-}=
\oneht ({{\bar q}_{\sss 1}}{q_{\sss 2}}+{{\bar p}_{\sss 1}}{p_{\sss 2}})-
\oneht ({{\bar q}_{\sss 1}}{p_{\sss 2}}+{{\bar p}_{\sss 1}}{q_{\sss 2}})\cr
&\quad\qquad\qquad\qquad(2,2)\qquad (1,1\oplus 3)\oplus (1\oplus 3,1)\cr
{\ket{\rm II}}&\sim
{{\bar\phi}^{\sss 1}_+}{\phi^{\sss 2}_+}=
\oneht ({{\bar q}_{\sss 1}}{q_{\sss 2}}+{{\bar p}_{\sss 1}}{p_{\sss 2}})+
\oneht ({{\bar q}_{\sss 1}}{p_{\sss 2}}+{{\bar p}_{\sss 1}}{q_{\sss 2}})\cr
&\quad\qquad\qquad\qquad(2,2)\qquad (1,1\oplus 3)\oplus (1\oplus 3,1)\cr
{\ket{\rm III}}&\sim
{{\bar\phi}^{\sss 1}_+}{\phi^{\sss 2}_-}
+{{\bar\phi}^{\sss 1}_-}{\phi^{\sss 2}_+}=
{{\bar q}_{\sss 1}}{q_{\sss 2}}-{{\bar p}_{\sss 1}}{p_{\sss 2}}\cr
&\qquad\qquad\qquad\qquad\qquad(2,2)\cr
{\ket{\rm IV}}&\sim
{{\bar\phi}^{\sss 1}_+}{\phi^{\sss 2}_-}
-{{\bar\phi}^{\sss 1}_-}{\phi^{\sss 2}_+}=
{{\bar p}_{\sss 1}}{q_{\sss 2}}-{{\bar q}_{\sss 1}}{p_{\sss 2}}\cr
&\quad\qquad\qquad\qquad(1,1\oplus 3)\oplus (1\oplus 3,1),\cr}
\EQN phenwhat$$
where we have used chiral symmetry to identify the wavefunctions with
the physical states of the previous section. These states have charge
conjugation sign: $\pm\epsilon$ for ${\ket{\rm I}}$, ${\ket{\rm II}}$
and ${\ket{\rm III}}$, and $\mp\epsilon$ for ${\ket{\rm IV}}$, and are
invariant with respect to the permutation

$$
{q_{\sss 1}}\longleftrightarrow{p_{\sss 1}}\quad  
{q_{\sss 2}}\longleftrightarrow{p_{\sss 2}}.
\EQN condi1a$$
The permutation symmetry which we used to constrain the hadron
wavefunctions (see \Eq{matrix2}) implied equal weight of the $(2,2)$
and $(1,1\oplus 3)\oplus (1\oplus 3,1)$ representations.  From the
point of view of the composites, this is a consequence of the
permutation 

$$
{q_{\sss i}}\longleftrightarrow{p_{\sss i}}\quad\;  
{q_{\sss j}},{p_{\sss j}}\quad
{\tt fixed}\;\quad{i\neq j},
\EQN condi$$
which interchanges $(2,2)$ and $(1,1\oplus 3)\oplus (1\oplus 3,1)$
representations and therefore leaves ${\ket{\rm I}}$ and ${\ket{\rm
II}}$ invariant while interchanging ${\ket{\rm III}}$ and ${\ket{\rm
IV}}$. 

It is clear that the meson states contain multiplets (a) and (b)
found above on the basis of symmetry arguments, as they
must. Moreover, now we see that since a state that transforms like
${\ket{\rm I}}$ must be identified with a
${{\bar\phi}_-}{\phi_-}$ state, Goldstone modes must be bound states
of ${\phi_-}$ quarks, as was conjectured based on the existence of a
conserved chiral current in the broken phase.

%%%%%%%%%%%%%%%%%%%%%%%%%%%%%%%%%%%%%%%%%%%%%%%%%%%%%%%%%%%%%%%%%%%
\vskip0.1in
\noindent {\twelvepoint{\bf 6.\quad Matching and the Mass Matrix}}
\vskip0.1in

We can learn about the meson mass matrix by using a simple
matching contraint. The leading explicit chiral symmetry breaking
operator in chiral perturbation theory implies that
${M_\pi^2}\propto{m_q}$. Since in the effective theory the
Goldstone bosons are bound states of quarks like other hadrons, and
all mass terms are on the same footing (see \Eq{masseswsb}), we expect
the constituent quark mass matrix to map to the hadronic {\it
mass-squared} matrix. The matching constraint fixes the meson masses
to be of the form:

$$\eqalign{
&{M_{\sss {I}}^2}={B}(M_{-}+M_{-})=2{B}(M-{M_0}+{m_q})\cr
{M_{\sss {III}}^2}&={M_{\sss {IV}}^2}={B}(M_{-}+M_{+})=2{B}(M+{m_q})\cr
&{M_{\sss {II}}^2}={B}(M_{+}+M_{+})=2{B}(M+{M_0}+{m_q}),\cr}
\EQN phen1$$
with associated meson quark content given in \Eq{phenwhat} and we have
used \Eq{masseswsb}; $B$ is an undetermined constant, which in
principle can be different for each chiral representation.  This
result maps to \Eq{mes1} and \Eq{mes2} only if $B$ transforms like
$(2,2)$; that is ${B}\rightarrow L{B}{R^\dagger}=R{B}{L^\dagger}$.
This is in fact the case for the representation involving the pion;
the pions are ${\phi_-}$ bound states with $M$=$M_0$ which gives
${M_\pi^2}=2B{m_q}$, where one identifies $B=-\vev{{\bar
q}q}/2{f_\pi^2}$ in QCD at leading order in chiral perturbation
theory.  We then have the following identification:

$$\eqalign{
&{\mu^2}=2{B}\{M+{m_q}\} \in {{\hat M}_0^2} \cr
&\delta =2{B}{M_0} \in {{\hat M}_{\sss \vev{{\bar q}q}}^2},\cr}
\EQN macphen1$$
which is consistent with the spurion transformation properties of
\Eq{spurtran}.  Note that the explicit breaking effects due to
current quark masses are contained in ${{\hat M}_0^2}$. Our results
are consistent with \Eq{mes1} and
\Eq{mes2} because the same symmetries are at work. In fact, \Eq{phen1}
follows from assuming that the quarks are confined and that the
matching constraint, ${M_\pi^2}\propto{m_q}$, is satisfied. However,
the matching constraint is correct only to leading order in chiral
perturbation theory. For instance, it receives contributions of higher
order in $m_q$ from loop graphs involving pions.  Therefore,
consistent matching requires that the $\phi_+$ and $\phi_-$
constituent quarks be weakly interacting in the same sense that
low-energy pions are weakly interacting. This is precisely what one
expects of constituent quarks.

%%%%%%%%%%%%%%%%%%%%%%%%%%%%%%%%%%%%%%%%%%%%%%%%%%%%%%%%%%%%%%%%%%%
\vskip0.1in
\noindent {\twelvepoint{\bf 7.\quad Phenomenology of the Ground State}}
\vskip0.1in

The lowest lying meson state must be a massless isovector, the pion,
which must be in a representation of type (a). Since the pion is a
Lorentz scalar, all states in this representation have
zero-helicity. In the case of zero-helicity there is conservation of
{\it normality}, $\eta\equiv P{(-1)^J}$, where $P$ is intrinsic parity
and $J$ is spin\ref{alg}. Since $\pi$ has $\eta$=$-1$, only states of
opposite normality communicate by single-pion emission and absorption.
Here we consider a well-known grouping.  The pion is joined by a
scalar $\epsilon$ ($\eta$=$+1$), and the helicity-$0$ components of
$\rho$ ($\eta$=$+1$) and ${a_{\sss 1}}$ ($\eta$=$-1$). These are
states with $GP{(-1)^J}$=$+1$ where $G$ is $G$-parity. Following
\Ref{alg} we identify
$\ket{\rm I}_a$=$\ket{\pi}_a$, $\ket{\rm{II}}_a$=$\ket{a_{\sss
1}}_a^{\sss (0)}$, $\ket{\rm{III}}$=$\ket{\epsilon}$ and
$\ket{\rm{IV}}_a$=$\ket{\rho}_a^{\sss (0)}$.  Here we review some
consequences of this grouping as well as constraints on the decay
constants that have not been considered previously using this
method. It is instructive to introduce an arbitrary mixing angle, $\phi$.
The pion representation is then given by:

\offparens
$$\eqalign{ 
&\ket{\pi}_a=-\cos\phi{\ket{2,2}_a}
+\sin\phi{\ket{A}_a}\cr
&\ket{a_{\sss 1}}_a^{\sss (0)}= \sin\phi{\ket{2,2}_a}
+\cos\phi{\ket{A}_a}\cr
&\ket{\epsilon}={\ket{2,2}_4}\qquad
\ket{\rho}_a^{\sss (0)}={\ket{V}_a}.\cr}
\EQN sfsr1$$ 
\autoparens
The superscript denotes helicity.  In order to learn about the
coupling of these states to the vector and axialvector currents in the
collinear frame, we can define the decay constants

\offparens
$$\eqalign{ 
&\bra{0}{{\cal A}_a}\ket{a_{\sss 1}}_b^{\sss (0)}={\delta_{ab}}f\cos\phi
\equiv{\delta_{ab}}{f_{a_{\sss 1}}}\cr
&\bra{0}{{\cal A}_a}\ket{\pi}_b={\delta_{ab}}f\sin\phi
\equiv{\delta_{ab}}{f_{\pi}}\cr
&\bra{0}{{\cal V}_a}\ket{\rho}_b^{\sss (0)}={\delta_{ab}}f
\equiv{\delta_{ab}}{f_{\rho}},\cr}
\EQN sfsr3$$ 
where

\offparens
$$
\bra{0}{{\cal V}_a}{\ket{V}_b}=
\bra{0}{{\cal A}_a}{\ket{A}_b}\equiv {\delta_{ab}}f .
\EQN sfsr3sdf$$ 
The usual definitions of the decay constants are 

\offparens
$$\eqalign{ 
&\bra{0}{A_{a\mu}}\ket{\pi}_b=
{\delta_{ab}}{f_{\pi}}{p_\mu}\cr
&\bra{0}{A_{a\mu}}\ket{a_{\sss 1}}_b^{\sss (\lambda )}=
{\delta_{ab}}{f_{a_{\sss 1}}}{M_{a_{\sss 1}}}
{\epsilon_\mu^{\sss (\lambda )}}\cr
&\bra{0}{V_{a\mu}}\ket{\rho}_b^{\sss (\lambda )}=
{\delta_{ab}}{f_{\rho}}{M_\rho}{\epsilon_\mu^{\sss (\lambda )}}\cr}
\EQN app1$$ 
where $\epsilon_\mu^{\sss (\lambda )}$ is the vector meson
polarization vector.  In the collinear frame
with $\lambda =0$ we recover
\Eq{sfsr3} from \Eq{app1} if we identify

\offparens
$$\eqalign{ 
&({{p^\alpha_0}-{p^\alpha_3}})
\bra{0}{{\cal A}^a}\ket{\alpha}\equiv\bra{0}{{A_0^a}-{A_3^a}}\ket{\alpha}\cr
&({{p^\alpha_0}-{p^\alpha_3}})
\bra{0}{{\cal V}^a}\ket{\alpha}\equiv\bra{0}{{V_0^a}-{V_3^a}}\ket{\alpha},\cr}
\EQN app4$$ 
where the 3-direction is the collinear direction of motion.  It
follows that $f_\pi$=$f_{\rho}\sin\phi$ and $f_{a_{\sss
1}}$=$f_{\rho}\cos\phi$.  Setting $M_\pi^2$=$0$ one obtains
${M_\rho^2}=\cos^2\phi{M_{a_{\sss 1}}^2}$. By considering matrix
elements of the pion transition operator, $X_a^\lambda$, one also
finds ${g_{\rho\pi\pi}^2}{f_\pi^2}={M_\rho^2}\sin^2\phi$\ref{alg}.
One can then obtain combinations of masses and decay constants that
are independent of the mixing angle. In particular it is clear that

\offparens
$$\EQNalign{ 
&\;\;{f_{a_{\sss 1}}^2}+{f_\pi^2}={f_\rho^2} \EQN sfsr4;a\cr
&{M_\rho^2}{f_\rho^2}={M_{a_{\sss 1}}^2}{f_{a_{\sss 1}}^2}\EQN sfsr4;b\cr}
$$\autoparens which is precisely the content of the first and second
spectral function sum rules\ref{sfsr}, respectively, evaluated in
resonance saturation approximation. This is not so surprising since in
both cases one can argue that chiral symmetry is the significant
input. Note that the approach used here is completely distinct from
the spectral function sum rule derivation which follows from
constraints on off-shell asymptotic behavior\ref{sfsr}.  Here we also
find other relations that are independent of mixing angle such as
${g_{\rho\pi\pi}^2}{f_\rho^2}={M_\rho^2}$.  Of course, understanding
of why the chiral multiplet takes its specific form requires the
$\discr$ permutation symmetry, or equivalently, the contraint
\Eq{mmconst2}.

The $\discr$ invariance implies that the irreducible chiral
representations must enter with equal weight and so $\pi$ and $a_{\sss
1}^{\sss (0)}$ form an $\discr$ doublet,
$\cos\phi$=$\sin\phi$=$1/\sqrt{2}$, and we obtain the familiar
KSRF relations\ref{ksrf}

\offparens
$$\eqalign{ 
&2{g_{\rho\pi\pi}^2}{f_\pi^2}={M_\rho^2}\cr
&2{f_\pi^2}{g_{\rho\pi\pi}}={M_\rho}{f_{\rho}},\cr}
\EQN sfsr6$$ 
which are remarkably well satisfied
experimentally\ref{pdg}\ref{alg}\ref{sfsr}.
\Eq{sfsr6} in turn implies ${f_{a_{\sss 1}}^2}={f_\pi^2}$ and
$2{M_\rho^2}={M_{a_{\sss 1}}^2}$.  We also obtain
${M_\epsilon^2}={M_\rho^2}$, and so the $\rho$ in the $\discr$
invariant four-flavor effective theory must be degenerate with a
scalar.  We emphasize that these relations, which have been derived
previously on the basis of asymptotic constraints\ref{alg}, are exact
consequences of $\discr$ in the four-flavor effective theory.

Note that the matching constraints of the previous section give the
mass-squared matrix of the pion ``quartet'' in terms of the
fundamental parameters of the theory. So, for instance, besides
${M_\pi^2}=2B{m_q}$, one also has ${M_\rho^2}=2B({M_0}+{m_q})$.
Therefore, in the four-flavor effective theory it is clear that as
$B\rightarrow 0$, at least $\pi$, $\epsilon$ {\it and} $\rho$
and ${a_{\sss 1}}$ become massless. So in the event of a second-order
phase transition, say at finite-temperature, one expects not the usual
$4$ of $O(4)$ sigma model scenario, but rather a new universality
class based on the reducible ``quartet'' representation found above
which in $O(4)$ notation corresponds to $4\oplus 6$\ref{unbreak}.

Here we have concentrated on the ground state.  The strong pion
transitions of heavy mesons have also been studied using the algebraic
sum rules\ref{bean}. There the sum rules have a great deal of
predictive power because heavy quark symmetry provides additional
constraints on the mass-squared matrix. We hope that flavor doubling
will provide some understanding of the strong transitions and
mass-squared splittings of the baryons, and of the manner
in which the various helicities are related\ref{beane}.

%%%%%%%%%%%%%%%%%%%%%%%%%%%%%%%%%%%%%%%%%%%%%%%%%%%%%%%%%%%%%%%%%%%
\vskip0.1in
\noindent {\twelvepoint{\bf 8.\quad Summary and Speculations}}
\vskip0.1in

In this paper we considered the possibility that QCD with ${{N}}$
flavors has a low-energy description with $2{{N}}$ flavors. We were
motivated by regularity in the hadron spectrum that is not explained
by QCD symmetries, particularly by asymptotic constraints in
pion-hadron scattering that can be expressed in algebraic form.  We
constructed a free theory of $2{{{N}}}$ quark flavors arranged into a
pair of vectors in the fundamental representation of $SU({{{N}}})$.
The free theory has a $U(2{{{N}}})$ invariance. However, we showed
that a $U(1)_A$ violating 't Hooft operator explicitly breaks
$U(2{{{N}}})$ to $SU({{{N}}})_L\times SU({{{N}}})_R\times U(1)_B$.  If
${\bar\theta}=0$ or $\pi$, the 't Hooft operator leaves
unbroken a discrete subgroup of $U(2{{{N}}})$: the group $\discr$ of
permutations of two objects.  We then assumed the QCD pattern of
chiral symmetry breaking and showed that in the $2N$-flavor effective
theory it is possible to move from the free current quark theory to
the free constituent quark model. In the constituent picture we
identified a conserved chiral current in the presence of chiral
symmetry breaking operators, and in turn conjectured that the
Goldstone bosons in the theory are bound states of massless
constituent quarks.

The $2N$-flavor effective theory was then studied on the light-front,
where we showed that a generalized Adler-Weisberger sum rule and
several superconvergence relations are manifest properties.  We then
showed that one can use $\discr$ and chiral symmetry to find the
chiral representations filled out by mesons in the broken phase,
obtained previously by assuming soft asymptotic
behavior\ref{mended}. Hadrons were then constructed directly out of
constituent quarks and the conjecture that Goldstone bosons are bound
states of massless quarks was proved.  We considered the phenomenology
of the chiral representation involving the pion and showed that
predictions of $\discr$ give familiar results that are in good
agreement with experiment.

Of course we have left several important questions unaddressed.
Perhaps the most relevant question is whether it is possible that QCD
has a low-energy effective description with a new symmetry which is
not present in the QCD lagrangian.

One interesting possibility is that $\discr$ is a discrete gauge
symmetry\ref{krauss}.  If $\discr$ is a gauge symmetry, then the
$2N$-flavor effective theory has the same global symmetries as
$N$-flavor QCD.  In this case, $\discr$ seems to solve the strong CP
problem without conflicting with expectations that global symmetries
are sacred and should therefore be shared by different descriptions of
the same physics. This interpretation is consistent with the fact that
$\discr$ only seems to have consequences related to asymptotic
behavior of scattering amplitudes.  Although gauge symmetries are
redundancies, they do have algebraic consequences when married with
asymptotic constraints on scattering amplitudes. In this sense gauge
symmetries behave like spontaneously broken chiral
symmetries\ref{alg}. A nice example is that of the
Drell-Hearn-Gerasimov sum rule which can be expressed algebraically as
a statement of the (trivial) $U(1)$ algebra of electromagnetism (see
the fourth entry in
\Ref{alg} and also \Ref{mended}).

The $\discr$ symmetry can also be interepreted as an accidental global
symmetry which is relevant only at energy scales where the $2N$-flavor
effective theory becomes a good description. There is some precedent
for this sort of accidental symmetry.  There exist supersymmetric
gauge theories where the full global symmetry group is not visible at
the level of the perturbative definition of the theory, but only in
the infrared, where there is an ``accidentally'' enhanced global
symmetry\ref{sei}.

A more ambitious interpretation of our results is that QCD with
${{N}}$ flavors has a dual ``magnetic'' description with $2{{N}}$
flavors.  This interpretation is suggested by the presence of the
chiral invariant mass term, $M_0$. This mass scale could arise
naturally from the condensation of scalar fields, which would
transform in the adjoint representation of the gauge group. Since
confinement is dual to the Higgs mechanism\ref{mandel}, one might then
interpret the fields in the effective theory with $2{{{N}}}$ flavors
as monopoles of the QCD degrees of freedom and condensation of the
``magnetic'' adjoint scalars as confinement of the QCD ``electric''
degrees of freedom.  Chiral symmetry breaking would seem to require
that the dual theory be in the Higgs phase since $M_0$ plays a
fundamental role in establishing the existence of a conserved chiral
current in the broken phase. Evidently the dual theory would have to
be in the Higgs phase with confined quarks and the ``magnetic'' gauge
theory would have the same number of colors as QCD in order that
baryons in the two descriptions be constructed out of the same number
of quarks.  This duality conjecture is not strictly academic.  As
pointed out above, a very interesting example of matching of distinct
descriptions exists on the light-front. A two-flavor quark model with
spin-orbit couplings\ref{suss} gives results for the mesons identical
to those found here in the four-flavor effective theory without
spin-orbit couplings. In the two-flavor theory $\discr$ must be
imposed by hand as a specific choice of symmetry breaking operators.
It would be interesting to see if a similar mapping exists for the
baryons\ref{beane}.

%%%%%%%%%%%%%%%%%%%%%%%%%%%%%%%%%%%%%%%%%%%%%%%%%%%%%%%%%%%%%%%%
\showsectIDfalse 
\section{Acknowledgements} 
This work was supported by the U.S. Department of Energy (Grant
DE-FG05-90ER40592 at Duke and grant DE-FG02-93ER-40762 at Maryland). I
thank T.D.~Cohen, M.A.~Luty, M.~Malheiro and B.~M\"uller for valuable
conversations and criticism.
\vfill\eject % new page 
\nosechead{References}% % no section number
%\addTOC{1}{References}{\folio}% % add to contents
%\global\def\HeadText{{\tenit References}}% % running head text
\ListReferences \vfill\supereject \end